\pdfoutput=1
\documentclass[letterpaper,10pt]{article} 
\usepackage{opticameet3} %% use version 3 for proper copyright statement

\usepackage{cite}
\usepackage{amsmath,amssymb,amsfonts}
\usepackage{algorithmic}
\usepackage{graphicx}
\usepackage{textcomp}
\usepackage{xcolor}
\usepackage[linesnumbered,ruled,vlined]{algorithm2e}
\usepackage{svg}
\usepackage{longtable}
\usepackage{float}
\usepackage{caption}
\usepackage{subcaption}
\usepackage{multirow}
\usepackage[export]{adjustbox}
\usepackage{soul}

\SetCommentSty{mycommfont}

\SetKwInput{KwInput}{Input}                % Set the Input
\SetKwInput{KwOutput}{Output}              % set the Output
\SetKwFunction{KwFn}{Fn}                   % set the Function
\usepackage[colorlinks=true,bookmarks=false,citecolor=blue,urlcolor=blue]{hyperref} %pdflatex

\svgpath{./images/}

\begin{document}

\title{Real-time, low latency virtual DBA hypervisor for SLA-compliant multi-service operations over shared Passive Optical Networks}

% \author{Author name(s)}
% \address{Author affiliation and full address}
% \email{e-mail address}
%%Uncomment the following line to override copyright year from the default current year.
%\copyrightyear{2022}

\vspace{-6mm}
\author{Arijeet Ganguli, Frank Slyne and Marco Ruffini}

\address{CONNECT Center, School of Computer Science and Statistics, Trinity College Dublin, Ireland}

\email{gangulia@tcd.ie, slynef@tcd.ie, marco.ruffini@tcd.ie} 

\vspace{-6mm}

\begin{abstract}
We present a heuristic algorithm for a PON upstream scheduling hypervisor, supporting low latency services with strict service-level agreement. The algorithm achieves near-optimal performance while running in only 3.5 $\mu s$, thus operating in real-time.
\end{abstract}
\vspace{-1mm}

\section{Introduction}
\vspace{-1mm}
Since the last decade, networks have started to migrate from closed, monolithic architectures towards open and disaggregated systems. Software Defined Networking (SDN) and Network Function Virtualization (NFV) have played a fundamental role in providing open interfaces for programmability and integration, adding flexibility in service development and provisioning, and bringing down costs by commoditising much of the networking and computing equipment. 
Passive Optical Networks (PONs) have experienced a similar transition, with early implementations, like the SDN Enabled Broadband Architecture (SEBA) \cite{seba}, enabling virtualisation of control and management planes. However, as PONs are being considered to offer services that go beyond residential broadband, especially supporting 5G and future 6G base stations \cite{convergence}, and Multi-access Edge Computing (MEC) \cite{MEC}, deep virtualisation, which enables control of scheduling algorithms, becomes important to meet these new performance targets.
A prominent example is the Flexible Access System Architecture (FASA) \cite{fasa}, which enables virtualisation and dynamic update of scheduling algorithms, which can be matched to the performance required by specific services. In \cite{vdba} we proposed a virtual Dynamic Bandwidth Allocation (vDBA) architecture for deep PON virtualisation, where independent Virtual Network Operators (VNOs) are able to run multiple different upstream scheduling algorithms (DBAs) in a shared PON. This enables both a multi-tenant and multi-service approach to passive optical networks. 

The approach consists on multiple VNOs running different schedulers in parallel, each of them proposing a virtual Bandwidth Map (vBMap) (which allocates upstream transmission slots to a group of Optical Network Units - ONUs). The key element is the scheduling hypervisor (which we also call merging engine), which collects all such virtual bandwidth maps to create a single physical bandwidth map that is transmitted to all ONUs (thus providing a solution that is compatible with typical PON standards).

The performance of the hypervisor determines the quality of service allocated to each VNO. Since the virtual BMaps from the different VNOs are independent, their proposed allocation will in general generate collisions (i.e., two independent BMaps propose scheduling of upstream resource over the same time sots). These needs to be resolved by the hypervisor, which has thus the ability to decide which allocations have to be rejected or delayed, when constructing the final bandwidth Map. An ideal hypervisor design is one that can make decisions based on specific Service Level Agreements (SLAs), so that it minimises the probability of breaching SLAs (which is key for supporting 5G and future 6G services). %In [] we have showed that type of optimisation requires the use of stateful algorithms, which make use of knowledge of previous allocations, to assure that a certain flow is not delayed by more than a target time us for a given percentage of time (for example, the vDBA method might not increase the delay of a specific flow by more than 10 $\mu s$ for 99\% of its time). 
The use of stateful algorithms, which take into consideration the history of a service flow when making scheduling prioritisation decisions, is preferred to stateless algorithms \cite{stateful}. This is because a stateful algorithm can prioritise flows depending on how close they are to breaching their specific SLA target. It should be noticed that this optimisation approach is not possible through classical queue management systems, where priority is allocated through stateless decisions, packet by packet.
The drawback of stateful algorithms is that they require computation times that can be incompatible with the short duration of PON frames, thus it is difficult to use them in practice on real systems. %In order to minimise the impact of adding this functionality to a PON, the hypervisor operation should be short compared to the PON frame time. %Thus we would expect this to contribute less than 10\% of the frame duration (i.e, \leq $12\mu s$). 

In this work we develop a heuristic stateful algorithm for the hypervisor that provides close to optimal performance, while running in real time. The algorithm is compared to an optimal MILP model to provide an upper bound on performance. We also provide comparison with a stateless algorithm based on simple packet priority rules as baseline reference for execution time. Our results show that the proposed heuristic can provide near optimal performance, with a run time of only 3.5$\mu s$, which is comparable to the run time of the stateless algorithm.
\vspace{-2mm}
\section{SLA Stateful DBA Hypervisor}  
\vspace{-2mm}
%The merging engine captures the various allocation reports from the VNOs in the form of respective BMaps and runs a scheduler that merges the BMaps and returns the merged BMap back to the VNOs to schedule their transmission accordingly. 
The aim of the stateful hypervisor is to minimise the probability of SLA breach for each individual service (or flow). An SLA breach occurs when a given flow accumulates a number of delayed upstream slots that is above its target SLA threshold. For example for an SLA with maximum upstream packet delay of 25 $\mu s$ with 99\% compliance, every time an upstream slot is delayed by more than 25 $\mu s$ with respect to the requested time slot in the virtual BMap, we increment a counter. If the counter goes above the non-compliance rate (in this case 100-99=1\%), calculated over a frame (i.e., of 125 $\mu s$ duration), then we consider that an SLA breach has occurred. The key idea of our real-time heuristic algorithm, reported in Fig. \ref{fig:pseudocode}, is thus to keep track of the history of the counter (through a flow-breach table shown in Fig \ref{fig:structure}) for each flow, and prioritise the flows based on the number of previous breaches. This allows to deal effectively with multiple flows even when they have different SLA targets (i.e., target latency and compliance rate).
The key data structure is a flow-breach likelihood table that keeps track of how far each traffic flow is from breaching its SLA (i.e., going above non-compliance rate). With reference to Fig. \ref{fig:pseudocode}, the heuristic first calculates the allocation maxtime which is the latest time an allocation can be scheduled within its latency target (code lines 1 to 3). Secondly, it starts allocating slots to the various allocations according to the time assigned by their originating virtual BMaps (code lines 5 to 6). Thirdly, it resolves collisions (lines 8 to 21) by allocating slots first in increasing order of non-compliance rate (line 18), then increasing order of their maxtimes (line 19) and finally increasing order of their sizes (line 20). Lines 22 to 26 initialise the SLA table shown in Fig. \ref{fig:structure}. Finally, the heuristic recalculates the non-compliance rate of all the flows and updates the flow-breach table for scheduling of the allocations in the next time frame (lines 27 to 32). %The likelihood is calculated as follows. The initial likelihood is the most negative value a flow can achieve when no packet level latency breaches has occured e.g if an SLA requires 95\% of its packets be scheduled within certain latency requirements, the negative likelihood is calculated as $100 - 95 = 5\%$. With every packet level breach, the likelihood approaches $0$ and a positive likelihood implies a flow-breach has occured. 
\begin{figure}
\hspace{-6mm}
\begin{minipage}[]{0.43\textwidth}
\vspace{-8mm}
\centering\includesvg[width=75mm,height=5cm]{OLT}
\caption{Stateful DBA Hypervisor}\label{fig:structure}
\end{minipage}
\hspace{2mm}
\begin{minipage}{0.57\textwidth}
\vspace{-8mm}
\centering\includesvg[width=95mm, height=14cm]{pseudocode}
\vspace{-6mm}\hspace{1cm}
\caption{Pseudocode of the heuristic stateful algorithm}
\label{fig:pseudocode}
\end{minipage}
\vspace{-8mm}
\end{figure}
%\vspace{-10mm}

Our heuristic is compared to a Mixed Integer Linear Programming (MILP) formulation, used as an upper bound for performance (but unable to run in real-time). The MILP model is shown in (Fig. \ref{fig:milp}): equations (1)-(4) calculate, respectively, the maximum delay of any given allocation to remain with the target threshold, the status of packet level breaches of the allocations after scheduling, the fraction of allocations in a flow that breached packet-level latency and the flow-level SLA breach status after scheduling. The objective (5) aims to minimize the overall flow-level SLA breaches defined in (4), by optimally allocating slots to the various virtual BMaps requests. Equations (6)-(8) puts constraints on the maximum possible size of the final merged BMap, slot allocation uniqueness (non-overlap) and maximum and minimum slot values.
     
\vspace{-2mm}
\section{Experiment and Performance Evaluation}
\vspace{-2mm}
We have carried out our experiments by feeding BMaps from different VNOs to our real-time hypervisor, running on an AMD-Ryzen7 4K Series Processor. The input allocation load on the shared PON was considered for 20\%, 50\% and 90\% of the total upstream capacity. For each of this allocation loads, we then varied the percentage allocated to SLA-driven flows from 10\% to 90\% of the total load (the remaining part is allocated to best effort flows). The comparative MILP implementation was executed on a CPLEX solver from IBM ILOG, which is a high performance solver for Linear Programming (LP), Mixed Integer Programming (MIP) and Quadratic Programming (QP) problems. 

The parameters for our experiment are as follows: we consider 5 VNOs (each one generating a virtual BMap every frame) and two types of SLAs (type-1: 95\% compliance for latency target of 12.5 $\mu s$, type-2: 90\% compliance for latency target of 25 $\mu s$). Each bandwidth map has a uniformly distributed set of allocations, and each individual burst allocation can be of 1.3KB, 4.7KB and 9.5KB, in average representing, respectively, about 1\%, 3\% and 6\% of the total frame size (the same ONU is also allowed to provide multiple burst per frame). An empty time slot of around 0.1 $\mu s$, is introduced between allocation to account for guard time between upstream transmissions. We also report the performance of a stateless algorithm based on simple prioritisation of one of the SLAs over the other, previously considered in \cite{stateful}.
%The upstream frame is split into 1152 allocation units or 125 $\mu s$. One unit is the smallest upstream grant that can be allocated to one ONU and for a 10G PON it has a size of 135 bytes and approximate duration of 0.11 $\mu s$[5]. In our scenario, three different average sizes are considered for the grants of 10, 35 and 70 units, measuring respectively, 1.3KB, 4.7KB and 9.5KB. The MIP implementation is run on a CPLEX solver from IBM ILOG which is a high performance solver for Linear Programming (LP), Mixed Integer Programming (MIP) and Quadratic Programming (QP) problems. 
The experiment was run for 1000 time frames (each of 125 $\mu s$) and the average run-time of the algorithm was recorded, together with the number of SLA breaches as a function of the allocation load. The plots report the ability to meet SLAs, depending on overall allocation load, percentage of load that is SLA oriented, and size of burst allocation (shown with different colors in the plots). 

\ul{The key part of the results are the conditions where the SLA are not breached (i.e., at the 100\% value of the y axis).} \ul{This information can be used by an operator to understand how much load from SLA-driven 5G services and other best effort services can be injected into a shared PON to achieve full SLA compliance.}

By comparing the plots in Fig. \ref{fig:heuristic} and Fig. \ref{fig:MILP}, we can see that our heuristic provides performance close to optimal, across all load scenarios, and most importantly provides similar compliance level to both types of SLAs, although they have different latency and compliance targets. For example, at 90\% overall load (lower plots), the heuristic can maintain the same 100\% compliance as the MILP (for up to 20\% of SLA-oriented flows in the system, shown in the x axis). For the 50\% load case (middle plots), we can see a slightly sub-optimal behaviour as our heuristic can meet SLAs up to a 40\% value in the x axis, while the MILP can reach 50\%.
We can also notice that in general the performance are lower for shorter burst, as this reduces the available capacity due to an increasing number of guard intervals (this reduction of performance for short bursts is typical in PON upstream allocations).
In addition, we can see that our heuristic performs much better than the baseline stateless algorithm reported in Fig. \ref{fig:stateless}, which, being based on simple packet-by-packet prioritisation, is only able to satisfy flows with one type of SLA (type-1), while it is totally non compliant for flows with SLA type-2.

\ul{We have also carried out run time profiling of the algorithms using the C-profiler gprof. This showed that every call to the hypervisor (i.e., for each physical BMap) is completed in average in 3.52 $\mu s$, which is only a small percentage (less than 3\%) of the 125 $\mu s$ frame duration.} It is also close to the run time of the stateless algorithm, running in 2.72 $\mu s$.

\begin{figure}[]
\begin{minipage}{1\textwidth}
\vspace{-15mm}
\begin{minipage}{0.5\textwidth}
\vspace{2mm}
\centering\includesvg[width=87mm,scale=0.4]{MIP_formulation}
\vspace{-7mm}
\caption{MILP Notations and Equations}
\label{fig:milp}
\end{minipage}
\vspace{-20mm}\hspace{0mm}
\begin{minipage}{0.6\textwidth}
\vspace{-5mm}\hspace{-14mm}
\centering\includesvg[width=93mm, scale=0.4]{stateless_sla}
\vspace{-58mm}\hspace{-4mm}
\caption{Stateless algorithm}
\label{fig:stateless}
\vspace{-5mm}
\end{minipage}
\end{minipage}
\begin{minipage}{1.0\textwidth}
\vspace{19mm}\hspace{-7mm}
\begin{minipage}{0.5\textwidth}
\centering\includesvg[width=94mm, scale=0.4]{stateful_sla}
\vspace{-75mm}\hspace{-2mm}
\caption{Stateful heuristic algorithm}
\label{fig:heuristic}
\end{minipage}
\vspace{25mm}\hspace{-1mm}
\begin{minipage}{0.5\textwidth}
\vspace{-1mm}\hspace{0mm}
\centering\includesvg[width=95mm, scale=0.4]{cplex_sla}
\vspace{-65mm}
\caption{MILP implementation}
\label{fig:MILP}
\end{minipage}
\end{minipage}
\vspace{-33mm}
\end{figure}
%\vspace{1cm}
\vspace{-2mm}
\section{Conclusions}
\vspace{-2mm}
In this work we presented a heuristic algorithm for a PON hypervisor, capable of satisfying multiple different SLAs, when scheduling upstream capacity in a multi-tenant PON infrastructure, thus supporting convergence of mobile and residential services. 
We have also shown the maximum percentage of SLA-oriented traffic that can be supported, for two sample SLA types, to satisfy requirements for low latency services with no SLA breach. This for example could support the Cooperative Transport Interface (CTI) over a shared PON, to transport fronthaul eCPRI signals for 5G base stations (and beyond). 
A key result is that the performance of our proposed heuristic is close to optimal and it is capable of running in real-time, with a run time less than 3\% of a frame duration.   
\vspace{-2mm}
\section{Acknowledgments}
\vspace{-2mm}
\small
Financial support from Science Foundation Ireland grants 12/RC/2276\_p2, 14/IA/2527 and 13/RC/2077\_p2 is acknowledged.

\end{document}